**Will pleural fluid affect surface wave speed measurements of the lung using lung ultrasound surface wave elastography: experimental and numerical studies on sponge phantom?**

Boran Zhou[1], Xiaoming Zhang[1]

[1]Department of Radiology, Mayo Clinic College of Medicine and Science, 200 1st St SW, Rochester, MN, 55905, USA.

Correspondence:

Xiaoming Zhang, PhD

Zhang.xiaoming@mayo.edu.

Department of Radiology, Mayo Clinic, 200 1st St SW, Rochester, MN, 55905, USA.

**Abstract**

Pleural effusion manifested as compression of pleural fluid on the lung parenchyma, contributing to hypoxemia. Medical procedure such as drainage of plural fluid releases this compression and increase the oxygenation. However, the effect of pleural effusion on the elasticity of lung parenchyma is unknown. By using the lung ultrasound surface wave elastography (LUSWE) and finite element method (FEM), the effect of pleural effusion on the elasticity of superficial lung parenchyma in terms of surface wave speed measurement was evaluated in a sponge phantom study. Different thickness of ultrasound transmission gel simulated as pleural fluid was inserted into a condom which was placed between the sponge and standoff pad. A mechanical shaker was used to generate vibration on the sponge phantom at different frequencies ranging from 100 to 300 Hz while ultrasound transducer was used to capture the motion for measurement of surface wave speed of the sponge. FEM was conducted based on the experimental setup and numerically assess the influence of pleural effusion on the surface wave speed of the sponge. We found the influence of thickness of ultrasound transmission gel was statistically insignificant on the surface wave speed of the sponge at 100 and 150 Hz both from experiments the FEM.

**Introduction**

Pleural effusion, frequently encountered in critically ill patients hospitalized in intensive care units (ICU), is the accumulation of excess fluid in the pleural cavity which is the fluid-filled space that surrounds the lungs. It induces restrictive syndromes and increases the intrapulmonary shunt by compressing the lung parenchyma. Study has showed that the pleural effusion may contribute to hypoxemia under mechanical ventilation [1]. In order to increase oxygen delivery and oxygen consumption, medical procedures such as drainage of pleural fluid is performed to increase functional residual capacity and improve oxygenation. Hemodynamic and pulmonary parameters, such as blood pressure, systemic vascular resistance, peak airway pressure, were collected before and after the fluid was drained [2]. However, no technique is able to evaluate the degree of restoration of the function of lung parenchyma.

In the ICU, the diagnosis of pleural effusion relies on the anteroposterior chest radiography obtained at bedside [3]. However, it exposes patients to a high dose of radiation. Pleural sonography is an alternative imaging modality. It is a highly portable and widely accepted diagnostic technique for identifying pleural disease [4, 5]. It permits imaging of pleural effusion and other pleural pathology. In addition, ultrasonography is able to guide thoracentesis for pleural interventions. Normal visceral and parietal pleura are apposed and 0.2-0.3 mm thick. Pleural effusions with parietal pleural thickness > 10 mm, and diaphragmatic thickness > 7 mm predict underlying malignancy with high specificity. It has been shown that a minimum pleural effusion depth of 1.2 cm between the visceral and parietal pleura has been recommended to perform diagnostic thoracentesis [6].

Thoracentesis is usually performed to relieve the compression of pleural effusion on the lung parenchyma. Few study evaluated the restoration of lung parenchyma after drainage of pleural fluid. Function of lung parenchyma is heavily dependent on its elasticity. In order to noninvasively evaluate the elasticity of lung parenchyma for patients with interstitial lung disease (ILD), we recently developed a lung ultrasound surface wave elastography (LUSWE) to measure the surface wave speed of lung which is correlated with lung elasticity [7-9]. No pleural effusion was observed for ILD patients so ultrasound propagation can penetrate the thoracic muscle and motion of lung surface can be captured by ultrasound imaging. For the patients with pleural effusion, the effect of pleural fluid on the surface wave speed of lung parenchyma in LUSWE is unknown. Therefore, there is a pressing need to develop a phantom model to systematically investigate the effect of pleural fluid on the measurements of LUSWE.

Wet foam dressing material has been used for lung ultrasound simulation models to teach novice physicians to perform lung ultrasound in clinical situations [10]. An economical sponge phantom was used for understanding and researching reverberation artifacts in lung ultrasound given its similar microstructure with lung parenchyma [11]. Moreover, with its availability, relevant phantom models for a systemic study of induced disease states, such as pulmonary edema, can be generated.

The objective of this study was to develop a phantom model for evaluating the effect of pleural fluid on surface wave speed in LUSWE. Ultrasound transmission gel to simulate pleural fluid was squeezed into a condom that was placed between the acoustic standoff pad and sponge phantom. In LUSWE, a shaker was used to generate a vibration on the surface of standoff pad, and the wave propagation on the surface of the sponge phantom was measured by using ultrasonic imaging and analysis. A FEM model was developed to simulate the wave propagation in the sponge phantom according to the experimental setup and compare with experimental measurements.

The rest of the paper is structured as follows: we describe the setup for the sponge phantom model in the Materials and Methods section; we present results that evaluate the effects of thickness of ultrasound transmission gel in the results section; we finalize the paper with discussion and conclusions.

**Materials and Methods**

The experimental setting consisted of the following parts: (1) household sponge (Ocelo utility sponge, 3M, St. Paul, MN); (2) ultrasound transmission gel (Aquasonic 100, Parker Laboratories Inc, Fairfield, NJ); (3) an acoustic standoff pad (Aquaflex; Parker Laboratories Inc, Fairfield, NJ). The acoustic standoff pad is made of a gel matrix free of air bubbles eliminating the air-filled space between the transducer and the sponge phantom. Without a standoff pad, extra-thoracic tissues will not be satisfactorily imaged. The fluid component of a pleural effusion may have echogenity which is characteristic of the presence of cellularity. Air bubbles within pleural fluid, which may occur with esophageal-pleural fistula or a gas-forming infection will exhibit multiple mobile echogenic foci within pleural fluid that represent air bubbles [12]. Ultrasound transmission gel has the similar echogenity as pleural fluid and can contain air bubbles in it. Sponge has been shown to have similar microstructure as lung parenchyma. Ultrasound transmission gel was squeezed into a condom that was placed between the standoff pad and sponge phantom. The thickness of gel in the condom was measured using ultrasound imaging and also taking pictures with a ruler placed aside as reference. The thickness of ultrasound transmission gel was varied at 4 levels: 0 (base), 2 mm (level 1), 7 mm (level 2) and 12 mm (level 3). A sinusoidal vibration signal of 0.1 s duration was generated by a function generator (Model 33120A, Agilent, Santa Clara, CA). The vibration signals were used at five frequencies of 100, 150, 200, 250, and 300 Hz [13]. The excitation signal at a frequency was amplified by an audio amplifier (Model D150A, Crown Audio Inc., Elkhart, IN). This signal then drove an electromagnetic shaker (Model: FG-142, Labworks Inc., Costa Mesa, CA 92626) mounted on a stand. The shaker applied a 0.1 s harmonic vibration on the surface of the acoustic standoff pad using an indenter with 3 mm diameter. 0.1 s is selected to exclude most of the reflections from the data collection window while keeping the detected wave as a continuous wave. The propagation of the vibration wave in the sponge was measured using a linear array transducer (L11-5v, Philips Healthcare, Andover, MA) transmitting at default 6.4 MHz center frequency mounted on the acoustic standoff pad. The transducer was connected to the ultrasound system (Vantage 1, Verasonics Inc, Kirkland, WA) (Fig. 1). The measurements were repeated three times at each frequency and each gel thickness.

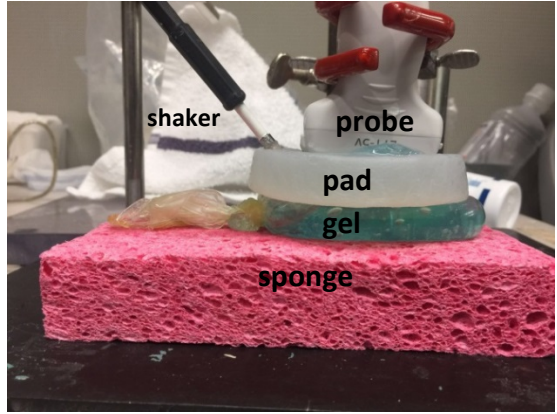

**Fig. 1.** Experimental setup of sponge phantom, transmission gel in a condom, and acoustic standoff pad.

Statistical analysis

An unpaired, two-tailed *t*-test of the differences in surface wave speed of the sponge phantom among different levels of gel thickness was conducted to compare sample means. Differences in mean values were considered significant at *p* < 0.05.

**Numerical modeling**

A FEM model was developed in ABAQUS (VERSION 6.14, 3DS Inc, Waltham, MA). The sponge phantom, acoustic standoff pad, and transmission gel in the condom were simulated as a 2D planar model of elastic medium (Fig. 2). Length and height of the acoustic standoff pad were 9 and 1.5 cm. Length and height of the sponge phantom were 12 and 2 cm. Ultrasound transmission gel thickness was predefined based on the measurements from experiments. The densities of the acoustic standoff pad and transmission gel were assumed to be 1000 kg m$^{-3}$. The structural constituent of the sponge phantom is cellulose, which has a density of 1500 kg m$^{-3}$. The sponge was modelled using a linear poro-viscoelastic model assuming the void ratio of the sponge is 0.7. The standoff pad was assumed an incompressible, linear elastic material [14]. To the best of our knowledge, viscoelastic properties of the sponge phantom have not been quantified. The values of the material parameters of the sponge phantom were estimated based on the experimental measurements in terms of surface wave speed of the sponge phantom at different vibration frequencies. Given the Voigt model has widely been used for characterizing the dynamic behavior of soft biological tissues, the Voigt model $\mu(t) = \mu_1 + \mu_2 \frac{\partial}{\partial t}$ is used in this study. The surface wave speed using this model was calculated as:

$$c_s = \frac{1}{1.05} \sqrt{\frac{2(\mu_1^2 + \omega^2 \mu_2^2)}{\rho \left( \mu_1 + \sqrt{\mu_1^2 + \omega^2 \mu_2^2} \right)}} \qquad (1)$$

where $c_s$, $\omega$, $\mu_1$ and $\mu_2$ are surface wave speed, angular frequency, shear elasticity, and shear viscosity, respectively. Given surface wave speeds of the sponge at five vibration frequencies, $\mu_1$ and $\mu_2$ were identified via least-square regression and used to calculate storage modulus $G_s$, loss modulus $G_l$, and long term modulus $G_{inf}$,

$$G_s = \mu_1 \qquad (2)$$

$$G_l = \omega \mu_2 \qquad (3)$$

$$G_{inf} = \mu_1 \qquad (4)$$

These quantities can then be implemented in the ABAQUS in the frequency domain with frequency ranging from 100 to 300 Hz at an interval of 50 Hz (Table 1).

**Table 1.** Material parameters of sponge phantom, acoustic standoff pad and ultrasound transmission gel.

| Materials | Sponge | Standoff pad | Transmission gel |
|---|---|---|---|
| E [kPa] | | 36.7 | |
| v | | 0.499 | |
| $\mu_1$ [kPa] | 6.83 | | 1.3 |
| $\mu_2$ [Pa·S] | 24 | | 24 |

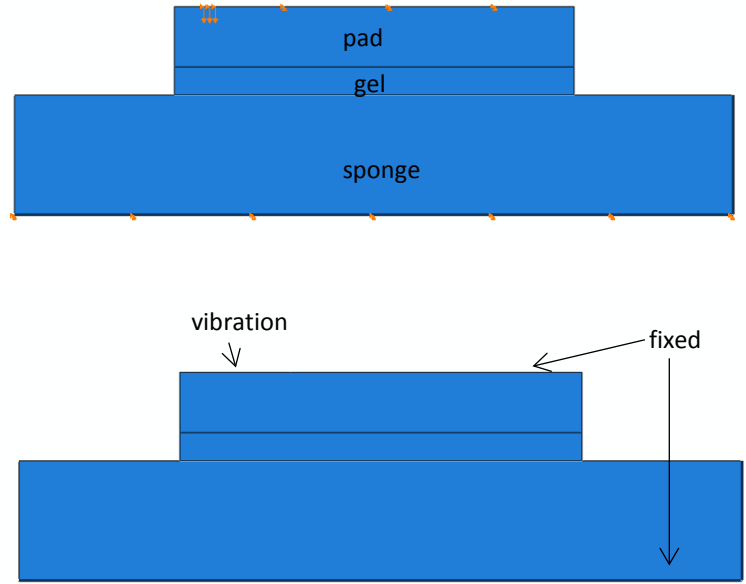

**Fig. 2.** Geometrical configuration of sponge phantom model.

The model was excited using a segment source on the top left surface of standoff pad and the displacement was applied in the vertical direction (Fig. 2). Harmonic excitations were performed at 100, 150, 200, 250, and 300 Hz with duration of 0.1 s. The central segment of the acoustic standoff top surface pad and bottom surface of sponge were fixed in the horizontal and vertical directions. The bottom boundary of the sponge phantom was attached to an infinite region to minimize the wave reflections [15].

The mesh of sponge, standoff pad, and ultrasound transmission gel in the condom were constructed using linear quadrilateral plane stress elements (type CPS4R) with size 1 mm x 1 mm, enhanced with hourglass control and reduced integration, to minimize shear locking and hourglass effects. The infinite region was meshed by infinite elements (type CINPE4) with size 1 mm x 1 mm. The dynamic responses of the sponge phantom model to the excitations were solved by the ABAQUS implicit dynamic solver with automatic step size control. Mesh convergence tests were performed so that further refining the mesh did not change the solution significantly.

**Results**

Detection of sponge motion is guided by ultrasound imaging. Sponge motion at a given location can be analyzed by cross-correlation analysis of the ultrasound tracking beam. In this study, eight locations on the surface of the sponge phantom over a length of approximately 8 mm were selected to measure sponge motion (Fig 3a). A high frame rate of 2000 frame/s is used to detect sponge motion in response to the vibration excitation at 100, 150, 200, 250, and 300 Hz. The surface wave speed is shown with 95% confidence interval, mean ± standard error (Fig. 3b). Fig. 6a shows the relationship between surface wave speed of sponge phantom and gel thickness. Surface wave speed (SWS) of the sponge phantom increased with frequency at each level of gel thickness, from 3.28 ± 0.08 m/s at 100 Hz to 7.43 ± 0.19 m/s at 300 Hz. At the same frequency, no statistically significant difference in surface wave speed of the sponge phantom was seen at different levels of gel thickness.

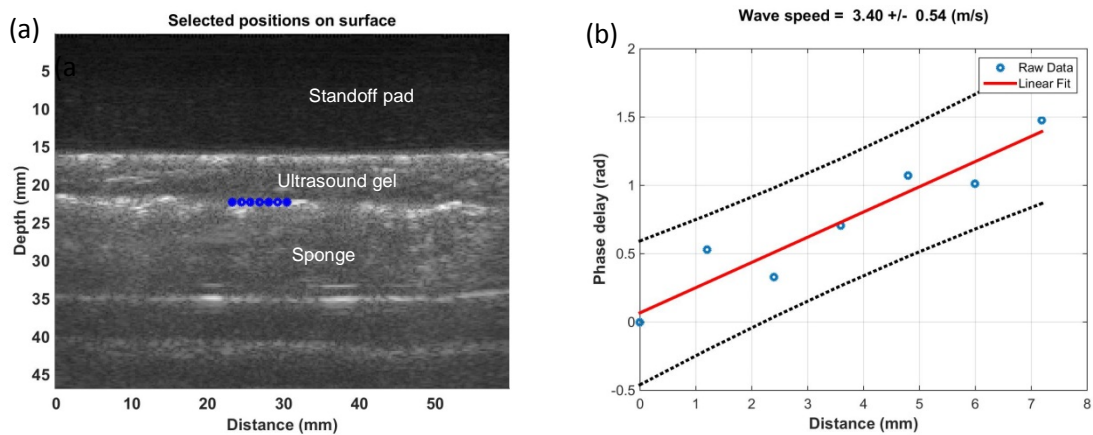

**Fig. 3.** (a) Representative B-mode image of sponge phantom model.

Eight locations on the surface of the sponge phantom were selected to measure the wave speed in the sponge by using the ultrasound tracking beam method. Blue dots indicate the points selected for measurement. (b) Representative phase delay-distance relationships of the sponge at frequency of 100 Hz. The wave phase change with position, in response to a 0.1 vibration was used to measure the wave speed.

FEM analysis of the sponge model was used to numerically investigate the effect of transmission gel thickness on the surface wave speed of the sponge phantom. Results of the finite element simulations are summarized in figure 4-6. As the boundary of the sponge was assigned infinite elements, there is no wave reflection in the boundary (Fig. 4). The temporal-spatial displacement field of a central segment of sponge was extracted to minimize the influence of boundary effects. The phase velocity can be measured using 2D Fourier transform using the fast Fourier transform (FFT) on the spatiotemporal motion data. The resulting Fourier distribution, or k-space, has one temporal frequency (F) axis and one spatial frequency (K) axis. 2D-FFT of the displacement versus time data was performed using

$$U_y(K,F) = \sum_{m=-\infty}^{+\infty}\sum_{n=-\infty}^{+\infty} u_y(x,t)e^{-j2\pi(Kmx+Fnt)} \tag{5}$$

where $u_y(x,t)$ is the motion of the sponge to the excitation as a function of distance from the excitation (x) and time (t). Here, K is the wave number and F is the temporal frequency of the wave. The coordinates of the k-space are the wave number (K) and the frequency (F) (Fig. 5) [16, 17]. For the harmonic wave case, a peak will occur at the excitation frequency, and the coordinates where the peak occurs can be used to determine the phase velocity using

$$c_s = \frac{f_p}{k_p} \tag{6}$$

where $c_s$ is the surface wave speed, $f_p$ is the peak temporal frequency and $k_p$ is the peak spatial frequency.

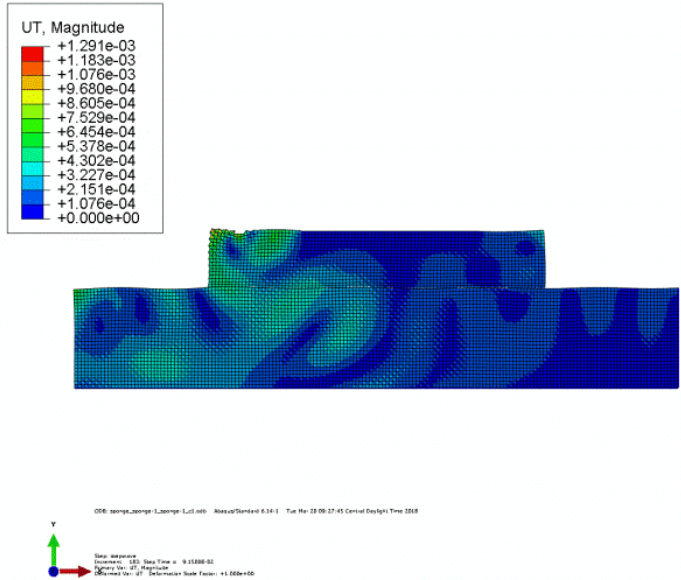

**Fig. 4.** Contour of translational displacement field of sponge phantom model due to harmonic vibration.

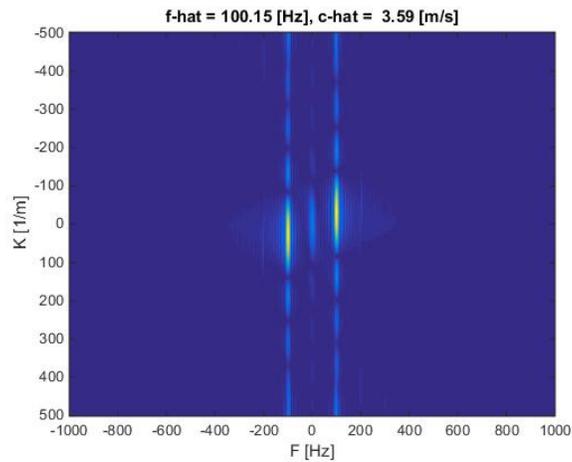

**Fig. 5.** Representative k-space from 2D FFT transformation of the sponge at 100 Hz excitation frequency.

The results from the numerical simulation showed that the surface wave speed (SWS) of the sponge phantom increased with excitation frequency at different levels of gel thickness (Fig. 6b). At 100 and 150 Hz, there was no significant difference in the surface wave speed of the sponge phantom at different levels of gel thickness. However, at 200, 250, and 300 Hz, there was increase in surface wave speed by 10% at ultrasound transmission gel thickness of 2 mm, 15% at 7 mm and 35% at 12 mm relative to the surface wave speed at base level of ultrasound gel thickness (Fig. 6b).

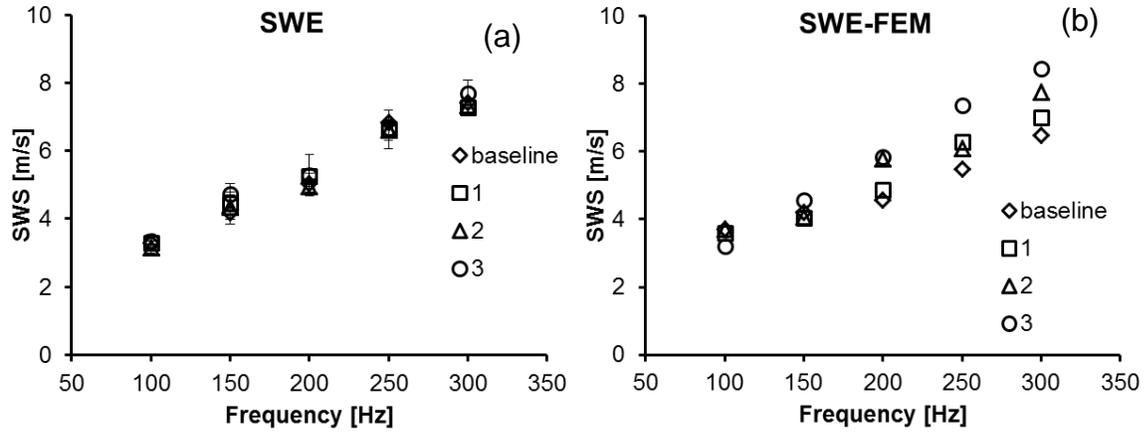

**Fig. 6.** (a) Surface wave speed – gel thickness levels of sponge phantom at multiple frequencies (100, 150, 200, 250, and 300 Hz) from 3 repetitive measurements. Error bars represent the standard deviation of 3 repetitive measurements. (b) Surface wave speed – gel thickness levels of sponge phantom at multiple frequencies (100, 150, 200, 250, and 300 Hz) from numerical simulations.

**Discussions**

The aim of this study was to develop a lung sponge phantom model to investigate the effect of pleural fluid on surface wave propagation in LUSWE. A sponge phantom, acoustic standoff pad, and ultrasound transmission gel were simulated as lung parenchyma, thoracic muscle, and pleural fluid. The level of transmission gel thickness was adjusted by squeezing different amount of transmission gel into a condom that was placed between the acoustic standoff pad and sponge phantom. At each level of gel thickness, a shaker was used to generate a harmonic mechanical vibration on the acoustic standoff pad at one frequency of 100, 150, 200, 250 and 300 Hz. The resulting wave propagation on the sponge surface was measured using an ultrasound transducer. A FEM model was developed according to the experimental setup to simulate the wave propagation on the surface of the sponge phantom. We found influence of ultrasound transmission gel thickness on the surface wave speed of the sponge insignificant at five different excitation frequencies from experiments. In FEM, at 100 and 150 Hz of excitation frequency, no significant influence of ultrasound transmission gel thickness on the surface wave speed of the sponge was observed yet at 200, 250 and 300 Hz, the surface wave speed on the sponge increased with thickness of ultrasound transmission gel in FEM simulation.

The dynamic response of the sponge phantom obtained in our study has qualitatively similar trends with that of lung parenchyma in vivo, as both exhibit increasing shear wave speed with excitation frequency [18]. The magnitude of surface wave speed of the sponge, which is an indicator of elasticity of the sponge, is higher than that of healthy subjects and similar with that of patients with interstitial lung disease [7, 8, 19]. The presented model is of particular relevance to human models and this technique may be adapted to assess patients with pleural effusion.

Lung ultrasonography is more sensitive, specific, and reproducible for diagnosing lung pathologies and can be considered an alternative to bedside check radiography and thoracic computed tomography [3]. It can be used efficiently to evaluate the lung since more than 70% of the lung can be imaged through intercostal spaces [20]. Lung ultrasonography is excellent for diagnosing pleural diseases, and it is especially useful in the emergency and critical care settings for the detection of pleural effusions or guidance of procedures such as thoracentesis [21, 22]. The volume of pleural fluid has been estimated via ultrasonography for mechanically ventilated patients or patients with thoracentesis [1, 23]. Characterization of elasticity of superficial lung parenchyma is critically important for evaluating the function of lung parenchyma [19, 24]. In LUSWE, the surface wave on the lung is safely generated by a local mechanical vibration on the chest. Diagnostic ultrasound is only used for detecting surface wave propagation on the lung. Hence, it is a noninvasive and safe technique for lung testing. With LUSWE, we plan to evaluate patients with pleural effusion via assessing elastic properties of superficial lung parenchyma prior and after drainage of pleural fluid.

Finite element modeling numerically evaluated the effects of ultrasound transmission gel on the wave propagation of the sponge surface. The ultrasound transmission gel used to simulate pleural fluid that occupies the pleural cavity is a gel-like substance. Given its mechanical behavior lies between Hookean solid and Newtonian fluid, the pleural fluid was modeled as a viscoelastic material with high viscosity. The infinite elements assigned on the boundary of muscle and lung were used to get rid of the wave reflection, and therefore, improve the accuracy of wave speed calculation. It showed that the wave speed increased with excitation frequency, which is in agreement with the obtained experimental measurements. In this study, the sponge was assumed as a linear, poro-viscoelastic material with a void ratio of 0.7 given its similar structure with lung parenchyma [25]. At higher frequencies, an increase in the surface wave speed of the sponge phantom at different levels of gel thickness could be due to a simplified assumption that the sponge phantom is a homogeneous material with void uniformly distributed in the sponge without air trapped within it. The experimentally measured shear wave speeds at five excitation frequencies of the sponge surface was used to identify its viscoelasticity and the Voigt model was used. Different viscoelastic models may yield different results, and their predictions will be studied in a separate future work.

We hold that this investigation represents a meaningful contribution to the collective knowledge of LUSWE, but acknowledge that there are limitations to the study that merit further consideration. Intercostal muscle was simulated using an acoustic standoff pad in this study. Structure and mechanical properties of intercostal muscle is more complicated than an acoustic standoff pad. The measured surface wave speed on the lung should have some contributions from the intercostal muscle. Lung sliding has been simulated using wet foam as lung parenchyma and hand as thoracic wall [10, 26]. Further improvement may use hand as thoracic wall or develop an ex-vivo animal model to investigate the effect of pleural fluid on the wave propagation of lung parenchyma in LUSWE.

**Conclusion**

In summary, the present manuscript develops a lung sponge phantom model to integrate experimental measurements and numerical simulation to characterize the effect of pleural fluid on the dynamic response of the sponge phantom in terms of surface wave speed. Both the experiments and the FEM analyses showed that ultrasound transmission gel thickness has an insignificant effect on the surface wave speed of the sponge phantom at excitation frequencies of 100 and 150 Hz. The lung sponge phantom model may be useful for further developing *ex vivo* animal lung models and *in vivo* human lung studies.

**Acknowledgements**

This study is supported by NIH R01HL125234 from the National Heart, Lung and Blood Institute. We would like to thank Mrs. Jennifer Poston for editing this manuscript.

**Conflict of Interest:** The authors declare that they have no conflict of interest.

**References**

1. Balik, M., et al., *Ultrasound estimation of volume of pleural fluid in mechanically ventilated patients.* Intensive care medicine, 2006. **32**(2): p. 318.

2. Ahmed, S.H., et al., *Hemodynamic and Pulmonary Changes after Drainage of Significant Pleural Effusions in Critically Ill, Mechanically Ventilated Surgical Patients.* Journal of Trauma and Acute Care Surgery, 2004. **57**(6): p. 1184-1188.

3. Lichtenstein, D., et al., *Comparative diagnostic performances of auscultation, chest radiography, and lung ultrasonography in acute respiratory distress syndrome.* Anesthesiology: The Journal of the American Society of Anesthesiologists, 2004. **100**(1): p. 9-15.

4. Remérand, F., et al., *Multiplane ultrasound approach to quantify pleural effusion at the bedside.* Intensive care medicine, 2010. **36**(4): p. 656-664.

5. Mayo, P.H. and P. Doelken, *Pleural ultrasonography.* Clin Chest Med, 2006. **27**(2): p. 215-27.

6. Soni, N.J., et al., *Ultrasound in the Diagnosis & Management of Pleural Effusions.* Journal of hospital medicine, 2015. **10**(12): p. 811-816.

7. Zhang, X., et al., *Lung ultrasound surface wave elastography: a pilot clinical study.* IEEE transactions on ultrasonics, ferroelectrics, and frequency control, 2017. **64**(9): p. 1298-1304.

8. Kalra, S., et al. *Lung ultrasound surface wave elastography-preliminary measurements in patients with interstitial lung diseases*. in *Respirology*. 2017. Wiley 111 River St, Hoboken 07030-5774, NJ USA.

9. Zhang, X., et al., *An ultrasound surface wave elastography technique for noninvasive measurement of surface lung tissue.* The Journal of the Acoustical Society of America, 2017. **141**(5): p. 3721-3721.

10. Lee, K.-H., et al., *Evaluation of a novel simulation method of teaching B-lines: hand ultrasound with a wet foam dressing material.* Clinical and experimental emergency medicine, 2015. **2**(2): p. 89.

11. Blüthgen, C., et al., *Economical Sponge Phantom for Teaching, Understanding, and Researching A‐and B‐Line Reverberation Artifacts in Lung Ultrasound.* Journal of Ultrasound in Medicine, 2017. **36**(10): p. 2133-2142.

12. Cardenas-Garcia, J., P.H. Mayo, and E. Folch, *Ultrasonographic evaluation of the pleura.* Pleura, 2015. **2**: p. 2373997515610270.

13. Kubo, K., et al., *The quantitative evaluation of the relationship between the forces applied to the palm and carpal tunnel pressure.* Journal of Biomechanics.

14. Chen, L., et al., *In Vivo Estimation of Perineal Body Properties Using Ultrasound Quasistatic Elastography in Nulliparous Women.* Journal of biomechanics, 2015. **48**(9): p. 1575-1579.

15. Zhou, B., A.J. Sit, and X. Zhang, *Noninvasive measurement of wave speed of porcine cornea in ex vivo porcine eyes for various intraocular pressures.* Ultrasonics, 2017. **81**(Supplement C): p. 86-92.

16. Bernal, M., et al., *Material property estimation for tubes and arteries using ultrasound radiation force and analysis of propagating modes.* The Journal of the Acoustical Society of America, 2011. **129**(3): p. 1344-1354.

17. Kubo, K., et al., *Ultrasound elastography for carpal tunnel pressure measurement: A cadaveric validation study.* J Orthop Res, 2017.

18. Xiaoming Zhang, B.Z., Thomas Osborn, Brian Bartholmai, James Greenleaf, Sanjay Kalra, *Assessment of interstitial lung disease using lung ultrasound surface wave elastography.* Ultrasonics Symposium (IUS), 2017 IEEE International, 2017: p. 1-4.

19. Osborn, T., et al. *A Non-Invasive Ultrasound Surface Wave Elastography Technique for Assessing Interstitial Lung Disease*. in *ARTHRITIS & RHEUMATOLOGY*. 2017. WILEY 111 RIVER ST, HOBOKEN 07030-5774, NJ USA.

20. Mayo, P.H. and P. Doelken, *Pleural ultrasonography.* Clinics in chest medicine, 2006. **27**(2): p. 215-227.

21. Mathis, G., *Atlas of chest sonography*. 2003: Springer Science & Business Media.

22. Hakimisefat, B. and P.H. Mayo, *Lung ultrasonography.* The Open Critical Care Medicine Journal, 2010. **3**(2): p. 21-25.

23. Vignon, P., et al., *Quantitative assessment of pleural effusion in critically ill patients by means of ultrasonography.* Critical care medicine, 2005. **33**(8): p. 1757-1763.

24. Xiaoming Zhang, B.Z., Sanjay Kalra, Brian Bartholmai, James Greenleaf, Thomas Osborn, *Quantitative assessment of scleroderma using ultrasound surface wave elastography.* Ultrasonics Symposium (IUS), 2017 IEEE International, 2017: p. 1-3.

25. Luo, H., I. Goldstein, and D. Udelson, *A three‐dimensional theoretical model of the relationship between cavernosal expandability and percent cavernosal smooth muscle.* The journal of sexual medicine, 2007. **4**(3): p. 644-655.

26. Shokoohi, H. and K. Boniface, *Hand Ultrasound: A High‐fidelity Simulation of Lung Sliding.* Academic Emergency Medicine, 2012. **19**(9).

1